\documentclass[aps, prd, twocolumn, superscriptaddress, showpacs]{revtex4}

\usepackage{amsmath, graphicx}

\def\beq{\begin{equation}}
\def\eeq{\end{equation}}
\def\d{\partial}

\newcommand{\Caltech}{\affiliation{Theoretical Astrophysics 350-17,
    California Institute of Technology, Pasadena, CA 91125}}

\newcommand{\Cornell}{\affiliation{Center for Radiophysics and Space
    Research, Cornell University, Ithaca, NY 14853}}

\begin{document}

\title{Spectral methods for the wave equation in second-order form}

\author{Nicholas~W.~Taylor} \Cornell \Caltech
\author{Lawrence~E.~Kidder} \Cornell
\author{Saul~A.~Teukolsky} \Cornell \Caltech

\date{\today}

\begin{abstract} 
Current spectral simulations of Einstein's equations require writing the equations in first-order form, potentially introducing instabilities and inefficiencies.  We present a new penalty method for pseudo-spectral evolutions of second order in space wave equations.  The penalties are constructed as functions of Legendre polynomials and are added to the equations of motion everywhere, not only on the boundaries.  Using energy methods, we prove semi-discrete stability of the new method for the scalar wave equation in flat space and show how it can be applied to the scalar wave on a curved background.  Numerical results demonstrating stability and convergence for multi-domain second-order scalar wave evolutions are also presented.  This work provides a foundation for treating Einstein's equations directly in second-order form by spectral methods.
\end{abstract}

% 04.25.D- Numerical relativity 
% 02.60.Cb Numerical simulation; solution of equations 
% 02.70.Jn Collocation methods 
% 02.70.Hm Spectral methods 

\maketitle

\section{Introduction}
Recent advances~\cite{Pretorius2005, Baker2006, Campanelli2006} in numerical simulations of black holes in general relativity have led to many interesting results.  Most of these simulations have been carried out with finite-difference methods.  However, the vacuum Einstein equations have mathematically smooth solutions (unless pathological coordinates are chosen).  Accordingly, one expects that spectral methods should be optimal in terms of efficiency and accuracy. 

Einstein's equations are a hyperbolic system involving second derivatives in space and time.  However, the numerical solution of hyperbolic systems using spectral methods is normally performed with a fully first-order formulation, even when the equations are naturally higher order.  Reducing the order of the equations is usually achieved by introducing new variables defined as first-order time or space derivatives.  The basic impetus for this first-order reduction is that there exists a well-established body of mathematical literature for first-order hyperbolic systems~\cite{GKO1995,Fornberg1998,Boyd2001}, which includes methods for analyzing the well-posedness of the equations and the proper way to impose stable boundary conditions in terms of characteristic variables.

The obvious disadvantage of the first-order reduction is the introduction of additional variables, whose definitions (at least for spatial derivatives) become constraints the solution must satisfy and thus new possible sources of instability in the system.  Furthermore, each new variable must be evolved, increasing the number of equations and the computational cost of the simulations.  In some cases, this can be a substantial increase.  

Successful simulations of Einstein's equations using spectral methods have thus far been implemented only as first-order reductions of the second-order system~\cite{Lindblom2006, Boyle2007}.  In the case of the generalized harmonic form of the equations, the reduction to first order in space proceeds by introducing 30 additional variables, more than doubling the number of equations and constraints in the system~\cite{Lindblom2006}.  These simulations typically require significant computational time, upwards of a hundred CPU-weeks for high resolution runs~\cite{Boyle2007}.  

A first order in time, second order in space system has the potential to reduce the constraint-violating instabilities and the computational expense of the simulations.  However, the mathematical knowledge underlying the proper formulation for such systems is much less developed.  Recently, Gundlach and Mart\'in-Garc\'ia have proposed and analyzed definitions of symmetric hyperbolicity for a general class of second order in space systems~\cite{GundlachMartinGarcia2004,GundlachMartinGarcia2006}.  They have also shown how one may define characteristic modes in the second-order system and thereby formulate stable boundary conditions at the continuum level.  

There still remains the problem of how to impose the boundary conditions in the discrete system (using spectral methods).  Even for the simplest representative hyperbolic system, the second order in space wave equation, naive attempts to impose boundary conditions in the same way as in a first-order formulation generally fail.  The difficulty is not due solely to the presence of second derivatives.  For example, methods exist for treating the second order spatial derivatives in the Navier-Stokes equations directly using spectral methods~\cite{HesthavenGottlieb1996,Hesthaven1997,Hesthaven1998}.  However, these techniques do not apply to the wave equation, as the characteristic structure is fundamentally different.  In this work, we present a new method for imposing boundary conditions in the second-order wave equation that is robust, stable, and convergent.

Since the generalized harmonic form of Einstein's equations appears as ten nonlinear coupled wave equations, this work provides a foundation for solving Einstein's equations directly in second-order form using spectral methods.  This application will appear in a subsequent paper~\cite{Taylor2010}.  It is likely that the work presented here will also allow other formulations of Einstein's equations, such as the BSSN (Baumgarte-Shapiro-Shibata-Nakamura) formulation, to be treated by spectral methods without reduction to first-order form.

In Section~\ref{Fosh1D} we review a typical spectral method for evolving the fully first-order form of the one-dimensional wave equation.  We review how boundary conditions can be imposed using penalty methods~\cite{FunaroGottlieb1988} and how stability of the system can be analyzed with energy methods~\cite{GKO1995, Fornberg1998}.  In Section~\ref{MainResult} we present the new penalty method for the one-dimensional second order in space wave equation and prove stability of the system using energy arguments.  In Section~\ref{3DSWF} we generalize the method to three dimensions, and in Section~\ref{3DSWC} we apply the method to the case of a scalar wave on a curved background.

\section{One-Dimensional Wave Equation} \label{1DSWF}

\noindent  We begin with the one-dimensional wave equation 
\beq
\ddot \psi = \psi^{\prime\prime}, \label{1Dwave}
\eeq
where $\psi=\psi(x,t)$.  Here, dots denote differentiation with respect to $t$, while primes denote differentiation with respect to $x$.  We will first review a typical first-order pseudo-spectral method for evolving this equation before discussing the second-order formulation.
  
\subsection{First-Order System} \label{Fosh1D}

\noindent The wave equation in Eq.~\eqref{1Dwave} reduces to first order by introducing the variables $\pi$ and $\phi,$ where 
\begin{flalign}
\pi &\equiv -\dot \psi, \label{PiDef1}\\
\phi &\equiv \psi^\prime. \label{1DVarphi}
\end{flalign}
The negative sign in the first equation is purely a matter of convention.  The first-order system is thus 
\begin{flalign}
\dot \psi &= -\pi,  \label{1DEq1}\\
\dot \pi &= -\phi^\prime, \label{1DEq2} \\
\dot \phi &= -\pi^\prime. \label{1DEq3}  
\end{flalign}
Equation~\eqref{1DEq1} is just the definition of $\pi$, while the definition of $\phi$ in Eq.~\eqref{1DVarphi} amounts to the addition of a constraint $\mathcal{C}=0$ to the system, where
\beq 
\mathcal{C} \equiv \psi^\prime - \phi.
\eeq
The characteristic variables $U$ and speeds $\lambda$ for this system (see e.g.\ Ref.~\cite{GKO1995}) are
\begin{flalign}
\quad \qquad \qquad U_\psi =&\, \psi, & \lambda &= 0, \qquad \qquad \quad \\
\quad \qquad \qquad U_\pm  =&\, \pi \pm n^x\phi, & \lambda &= \pm 1. \qquad \qquad \quad \label{UpmDef1D}
\end{flalign}
Here, $n^x$ is the unit outgoing normal vector to the boundary, which in one dimension is just $n^x = \pm 1$.  With this definition, $U_-$ is incoming ($\lambda < 0$) at each boundary.  

For a symmetric hyperbolic system on a domain $\Omega$ with boundary $\partial \Omega$, there exists a (not necessarily unique) conserved, positive definite energy 
\beq
E=\int_\Omega \! \epsilon \,dV, 
\eeq 
which is conserved in the sense that
\beq
\dot \epsilon = \d_iF^i.  \label{FluxDef}
\eeq 
Accordingly, the time derivative of the energy is given by the flux through the boundary,
\beq
\dot E = \int_{\d \Omega} F^n \,dA,
\eeq
where $F^n \equiv n_i F^i$.  Note that for general quasi-linear systems, the energy is only strictly conserved when coefficients in the equations are approximated as constant and lower-order terms are neglected.  

For the one-dimensional wave equation in Eqs.~\eqref{1DEq1}-\eqref{1DEq3}, the energy density is
\beq
\epsilon = \frac{1}{2}(\pi^2 + \phi^2). \label{1DEnergy}
\eeq
Using Eqs.~\eqref{FluxDef},~\eqref{1DEq2},~\eqref{1DEq3}, and~\eqref{UpmDef1D}, we get
\beq
F^x = -\pi\,\phi = \frac{n^x}{4} (U_-^2 - U_+^2).
\eeq
If we consider our domain to be the interval $[-1,1]$, then
\beq
\dot E = \frac{1}{4} \sum\limits_{x=\pm1} (U_-^2 - U_+^2). \label{1dflux}
\eeq
For well-posedness and stability, one requires that the growth of the energy be controlled by specifying boundary conditions for the positive terms in $\dot E$.  Therefore, a boundary condition must be supplied on the incoming mode $U_-$.  For example, with a homogeneous condition specifying $U_-=0$ (or more generally $U_-=\kappa\,U_+$ for $|\kappa|\le 1$), it follows that $\dot E \le 0$. Together with the positive definiteness of the energy, this ensures that the system is stable.  If instead the incoming mode is a prescribed function $U_-=f$, then we still have stability in the sense of controlling the energy with a bound that involves $f$.

The definition of energy given by Eq.~\eqref{1DEnergy} is not unique, but was motivated in part by a desire to obtain a sharp energy bound.  For example, we could have defined the energy density with a term $a^2\psi^2$ as 
\beq
\epsilon = \frac{1}{2}(a^2\,\psi^2 + \pi^2 + \phi^2).
\eeq
In this case, we would obtain the additional term in $\dot E$
\beq
-\int_\Omega a^2\,\psi \,\pi \,dV \le \frac{a}{2} \int_\Omega (a^2\psi^2 + \pi^2) \,dV,
\eeq
where the inequality follows from the relation $2\,uv \le u^2 + v^2$ for any (real) $u,v$.  We would thus arrive at the weaker estimate
\beq
\dot E \le \frac{1}{4} \sum\limits_{x=\pm1} (U_-^2 - U_+^2) + a\,E. \label{psi2flux}
\eeq
With a condition on the incoming mode $U_-$ to control the boundary term on the right-hand side, the system is still well-posed in this case~\cite{GKO1995}, but we are unable to prove stability in the sense that $\dot E \le 0$. 

In the semi-discrete problem, one considers the discretization in space but not time. An effective way to impose boundary conditions in the semi-discrete system is to add appropriate penalty terms to the equations on the boundaries.  Such a penalty method thus imposes conditions ``weakly''---that is, approximately, without completely replacing the equation of motion on the boundary~\cite{FunaroGottlieb1988}.  Heuristically, the rationale for this is that it is not necessary to enforce exact boundary conditions on approximate solutions.  All that is required is for the discrete solution to converge to the continuum solution with the correct boundary conditions as the resolution is increased.  We find that these methods generally yield superior accuracy and convergence while also providing a simple way to impose arbitrary boundary conditions.

The penalties are added to the equations on the boundary in the form $(U_-^\text{BC}-U_-)$, so that if the boundary condition $U_- = U_-^\text{BC}$ is satisfied then the penalties vanish.  The appropriate penalty (up to an overall coefficient) for each equation can be found by projecting the boundary conditions in terms of characteristic variables to fundamental variables~\cite{Bjorhus1995}.  In other words, we first transform to characteristic variables in the first-order system of Eqs.~\eqref{1DEq1}-\eqref{1DEq3} on the boundary and add penalties:
\begin{flalign}
\dot U_\psi &= -\frac{1}{2}(U_+ + U_-),  \\
\dot U_+ &= -n^xU_+^\prime,  \\
\dot U_- &= +n^xU_-^\prime + c\,(U_-^\text{BC} - U_-),
\end{flalign}
where $c$ is an undetermined constant.  Only the equation for $\dot U_-$ has a penalty term, since there is no boundary condition on $U_\psi$ or $U_+$.  We then transform back to fundamental variables to obtain the first-order equations with penalties:
\begin{flalign}
\dot \psi_i =&\, -\pi_i, \label{Fosh1} \\
\dot \pi_i =&\, -\phi_i^\prime + \frac{c}{2}\,(\delta_{i0}+\delta_{iN})(U_-^\text{BC} - U_-), \label{Fosh2} \\ 
\dot \phi_i =&\, -\pi_i^\prime - \frac{c}{2}\, n^x (\delta_{i0}+\delta_{iN})(U_-^\text{BC} - U_-). \label{Fosh3}
\end{flalign}
Here we have explicitly denoted grid values with a subscript $i$.  For a pseudo-spectral method, one chooses the nodes of a Gaussian quadrature rule as collocation points.  The $N+1$ grid points $x_i$ run from $x_0=-1$ to $x_N=+1$.  Differentiation is implemented by matrix multiplication, as in $\pi_i^\prime \equiv \sum_j D^{(1)}_{ij} \pi_j$, where $D^{(1)}_{ij}$ is the first-order differentiation matrix.  The Kronecker delta terms $\delta_{i0}+\delta_{iN}$ indicate that the penalties are applied only on the boundaries at $i=0,N$.  The penalty coefficients should satisfy $c \rightarrow \infty$ as $N \rightarrow \infty$, in order to ensure that the continuum equations and boundary conditions are recovered in this limit~\cite{FunaroGottlieb1988}. 

Suitable values for the penalty factor $c$ in Eqs.~\eqref{Fosh2}-\eqref{Fosh3} can be determined from a semi-discrete energy analysis, which we will now show.  For ease in obtaining analytical results, we choose \textit{Gauss-Legendre-Lobatto} collocation points (see Appendix~\ref{appGLLnotes} for details).  The basis functions for this choice are the Legendre polynomials $P_n(x)$ on $[-1,1]$.  We begin by writing the semi-discrete energy corresponding to Eq.~\eqref{1DEnergy}:
\beq
E = \frac{1}{2}\left[\langle \pi,\pi \rangle + \langle \phi,\phi \rangle\right], \label{FoshSDE}
\eeq
where $\langle \,\cdot\, , \,\cdot\, \rangle$ represents a discrete inner product, as in
\beq
\langle \pi, \pi \rangle \equiv \sum\limits_{i=0}^{N} \omega_i \,\pi_i^2.
\eeq
Here $\pi_i$ are the grid values of the function $\pi$, and $\omega_i$ are the quadrature weights (see Appendix~\ref{appGLLnotes}).  Taking the time derivative of the semi-discrete energy in Eq.~\eqref{FoshSDE}, we obtain
\beq
\begin{split}
\dot E =&\, -\pi_i \phi_i \big|_{i=0}^N + \frac{c}{2} \langle \pi, (\delta_{i0}+\delta_{iN})\delta U_- \rangle  \\
&\>- \frac{c}{2}\langle \phi,(\delta_{i0}+\delta_{iN})n^x\delta U_- \rangle, \label{FoshEeq1}
\end{split}
\eeq
where we have used summation by parts (the discrete analogue of integration by parts) in the first term and introduced the notation $\delta U_-\equiv U_-^\text{BC}-U_-$ in the penalty terms.  The first term in Eq.~\eqref{FoshEeq1} is similar to the continuum result:
\beq
-\pi_i \phi_i \big|_{i=0}^N = \frac{1}{4} \sum\limits_{i=0,N} (U_-^2 - U_+^2).
\eeq
Evaluating the discrete inner products in the last two terms in Eq.~\eqref{FoshEeq1} yields 
\beq
\dot E_\text{penalties} = \frac{c\,\omega}{2} \sum\limits_{i=0,N}  U_- \delta U_-,
\eeq
where we have written $\omega$ for the quadrature weight $\omega_0=\omega_N$ at $x=\pm1$.  Noting that 
\beq
U_-\delta U_- = \frac{1}{2}\big( U_-^\text{BC}{}^2 - U_-^2 - \delta U_-^2\big),
\eeq
we put things together to find
\beq
\dot E = \frac{1}{4}\sum\limits_{i=0,N} \big[(1-c\,\omega )U_-^2  -U_+^2 + c\,\omega\,( U_-^\text{BC}{}^2 - \delta U_-^2 ) \big]. \label{FoshEdot}
\eeq
The condition on the penalty factor $c$ for stability depends on the boundary condition we impose on $U_-^\text{BC}$.  Requiring $\dot E \le 0$, we find:
\begin{flalign}
U_-^\text{BC} &=  0 \qquad \>\>\, \Rightarrow   \quad c\ge  \frac{1}{\omega}, \\
U_-^\text{BC} &=  \kappa\,U_+  \quad \Rightarrow \quad \frac{1}{\omega\, \kappa^2} \ge  c \ge \frac{1}{\omega},
\end{flalign}
where $|\kappa| \le 1$.  The strictest condition is obtained by insisting that the energy be bounded by the continuum energy in Eq.~\eqref{1dflux} for arbitrary $U_-^\text{BC}$:
\beq
\dot E \le \dot E_\text{continuum} \> \iff \> c=\frac{1}{\omega}. \label{cCond}
\eeq

The situation is slightly different when considering the semi-discrete energy for a multi-domain problem.  For example, suppose we consider the interval $[-2,2]$ with an inner boundary at $x=0$.  The energy calculation up to Eq.~\eqref{FoshEdot} is identical on each subdomain.  The key difference is that now the incoming mode at the interface boundary is supplied by the adjacent subdomain.  If we denote the intervals $[-2,0]$ and $[0,2]$ with subscripts $1$ and $2$, respectively, then at $x=0$:
\begin{flalign}
U_{1-}^\text{BC} &= U_{2+}, \\
U_{2-}^\text{BC} &= U_{1+}.
\end{flalign}
The terms in $\dot E$ at $x=0$ are then:
\beq
\begin{split}
&(1-c\,\omega)U_{1-}^2 - (1-c\,\omega)U_{1+}^2 - c\, \omega \, \delta U_{1-}^2 \\
+ \,&(1-c\,\omega)U_{2-}^2 - (1-c\,\omega)U_{2+}^2 - c\, \omega \, \delta U_{2-}^2.
\end{split}
\eeq
This quadratic form is negative semi-definite if and only if $c=1/\omega$.  On a multi-domain problem, the value of $c$ at an internal boundary required for stability is therefore fixed, regardless of the boundary condition imposed at the external boundaries.  This analysis assumes that the penalties at the interface boundary enforce conditions only on the incoming modes.  It is also possible to penalize arbitrary combinations of the variables at interfaces and thereby to obtain different stability conditions (see e.g.\ Ref.~\cite{LehnerReulaTiglio2005}), but we do not consider this refinement here.

On an arbitrary domain, the definition of the discrete inner product is modified.  For instance, if we consider a one-dimensional domain $\Omega$ with a coordinate mapping $\mu:[-1,1] \rightarrow \Omega$, then the Jacobian of the mapping is inherited from the continuum inner product:
\beq
\langle f, g \rangle \equiv \sum\limits_{i=0}^N \omega_i\, f_i \,g_i\, \mu^\prime_i. \label{1dModIP}
\eeq
Since the penalty terms in Eq.~\eqref{FoshEeq1} contain Kronecker deltas that pick out specific terms from the sums, the values for $c$ we obtain would need to be modified by a Jacobian factor: $c \rightarrow c/\mu^\prime$. For simplicity, we will assume that the domain is the fundamental interval $[-1,1]$ unless otherwise stated, so that no Jacobians are needed.

Although we performed the semi-discrete energy analysis on Legendre grid points, this is not a limitation.  One could implement the system on Gauss-Chebyshev-Lobatto points using, for example, the Chebyshev-Legendre method~\cite{DonGottieb1994}.  With this method, the equations are implemented on a Chebyshev grid by interpolating the (Legendre-grid) penalty functions to the Chebyshev points.  In particular, a penalty that is applied only on the boundary of a Legendre grid as in Eqs.~\eqref{Fosh2}-\eqref{Fosh3} would in general be non-zero everywhere on a Chebyshev grid.  In practice, the system works well even without modification on a Chebyshev grid by simply using Eqs.~\eqref{Fosh1}-\eqref{Fosh3} as derived for a Legendre grid and letting the index $i$ represent the Chebyshev grid points.  Chebyshev stability of penalty methods is proved in Ref.~\cite{FunaroGottlieb1988} for simple cases, and proofs of Chebyshev stability for more general hyperbolic problems are reviewed in Ref.~\cite{GottliebHesthaven2001}.      

It is also worth noting that stability conditions derived from strict energy arguments can generally be relaxed to a degree.  The penalty factor $c$, which was found to be $1/\omega = N(N+1)/2$ for Legendre methods, can be optimized by trial and error to maximize efficiency and obtain the least restrictive \textit{Courant-Friedrichs-Lewy} (CFL) condition while maintaining stability.  This is discussed, for example, in Ref.~\cite{HesthavenGottlieb1996}. 

Stability of the \textit{fully} discrete problem can be explored by examining eigenvalues.  The entire system is written in a form suitable for passing to an explicit time-stepping algorithm, as in $\dot y = A y$, where the vector $y$ represents the grid values of all the fields.  The eigenvalues of the matrix $A$ can then be plotted in the complex plane and compared with the stability region of the time-stepper.  In general, positive real parts of the eigenvalues imply instability, while the spectral radius (maximum amplitude of eigenvalues) is inversely proportional to the maximum allowed time-step~\cite{GKO1995} (the exact relation depends on the time-stepping algorithm being used).  

\begin{figure*}[t!]
\centering
\begin{tabular}{cc}
\includegraphics[width=\columnwidth]{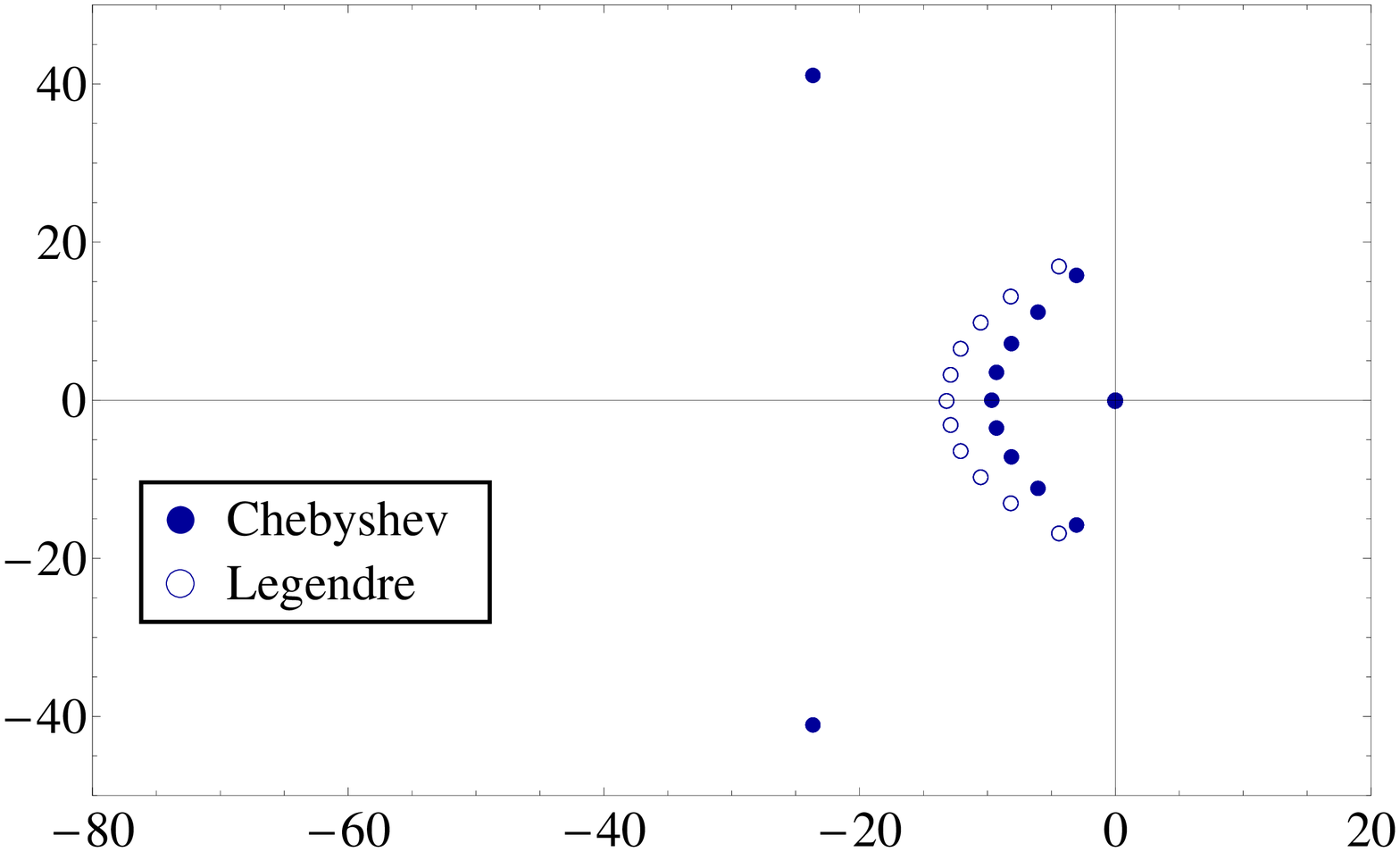} &
\includegraphics[width=\columnwidth]{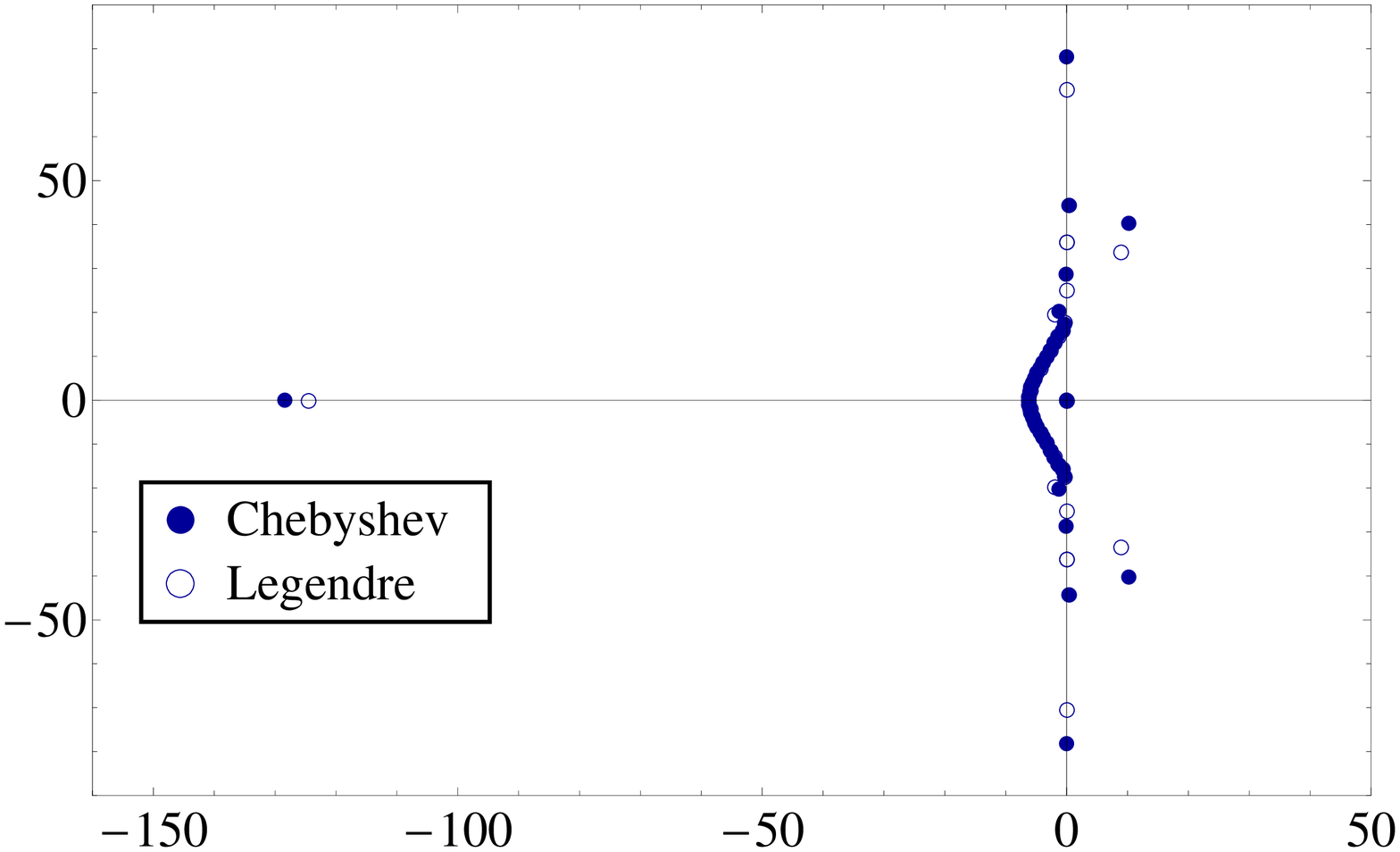}
\end{tabular}
\caption{Left: Eigenvalues in the complex plane of the first-order system, Eqs.~\eqref{Fosh1}-\eqref{Fosh3}. Right: Eigenvalues of a typical unstable second-order system, Eqs.~\eqref{Fotsos1}-\eqref{Fotsos2}.  Both plots are for a two-domain problem on [-1,1], with an inner boundary at $x=0$, penalty factors $c=N(N+1)/2$, outer boundary conditions $U_-^\text{BC}=0$, and $N+1=11$ grid points per subdomain. Results for Legendre- and Chebyshev-Lobatto grids are shown for comparison. \label{fg:evalues}}
\end{figure*}

A typical eigenspectrum for the system in Eqs.~\eqref{Fosh1}-\eqref{Fosh3} on two subdomains is shown in Fig.~\ref{fg:evalues}.  Curiously, the large amplitude conjugate pair of eigenvalues on Chebyshev points is absent on the Legendre grid.  This implies that there is a less restrictive CFL condition for the system on Legendre grid points, and this is indeed the case for this particular system.  As is discussed elsewhere, it is unlikely that this difference carries over to more general systems~\cite{Fornberg1998}.  For instance, we find no significant difference in time-stepping conditions on Chebyshev or Legendre grids for the \textit{three-dimensional} wave equation (in flat or curved space).  It is also worth noting that eigenvalue stability is insufficient to prove that the system is actually stable and convergent, but it is suggestive~\cite{Fornberg1998}.

%======================= FIGURE Fosh vs. bad Fotsos Eigenvalues  ========================
%=========================================================================

\subsection{Second Order in Space}

\noindent The first order in time, second order in space formulation of the one-dimensional wave equation in Eq.~\eqref{1Dwave} is
\begin{flalign}
\dot \psi =&\, -\pi, \\
\dot \pi =&\, -\psi^{\prime\prime}.
\end{flalign}
The characteristic variables are the same as those of the first-order reduction with the replacement $\phi \rightarrow \psi^\prime$:
\begin{flalign}
\quad \qquad \qquad U_\psi =&\, \psi, & \lambda &= 0, \qquad \qquad \quad \\
\quad \qquad \qquad U_\pm  =&\, \pi \pm n^x\psi^\prime, & \lambda &= \pm 1, \qquad \qquad \quad
\end{flalign}
The energy and flux are the same as well:
\beq
\epsilon = \frac{1}{2}(\pi^2 + \psi^\prime{}^2) \quad \Rightarrow \quad \dot E = \frac{1}{4} \sum\limits_{x=\pm1} (U_-^2 - U_+^2). \label{1dEcont}
\eeq
The difficulty arises in the semi-discrete second-order system when trying to find appropriate penalties by projecting from characteristic variables, as was done in the first-order case.  The boundary condition $\delta U_- \equiv U_-^\text{BC} - U_- =0$ is now a \textit{differential} as opposed to an \textit{algebraic} condition:
\beq
\pi - n^x \psi^\prime = U_-^\text{BC}.
\eeq
One therefore obtains a condition on $\psi^\prime$ at the boundary, but not on $\psi$ itself (there is no boundary condition on $U_\psi$).  The system one arrives at by naively following the same procedure as in the first-order case is
\begin{flalign}
\dot \psi_i =&\, -\pi_i,  \label{Fotsos1} \\
\dot \pi_i =&\, -\psi_i^{\prime\prime} + c\,(\delta_{i0}+\delta_{iN})\,\delta U_-. \label{Fotsos2}
\end{flalign}
We might try applying a penalty to Eq.~\eqref{Fotsos1} also:
\begin{flalign}
\dot \psi_i =&\, -\pi_i + c_1 (\delta_{i0}+\delta_{iN})\,\delta U_-, \label{Fotsos3} \\ \noalign{\vskip2pt}
\dot \pi_i =&\, -\psi_i^{\prime\prime} + c_2\,(\delta_{i0}+\delta_{iN})\,\delta U_-. \label{Fotsos4}
\end{flalign}
These equations are generally unstable, particularly when evolved on multiple subdomains with at least one interface boundary.  The error in $\psi$ tends to grow exponentially, ruining the evolutions within a few hundred crossing times.  The penalty factors $c_1, c_2$ can be fine-tuned by trial and error to obtain approximately stable evolutions in some cases, but not robustly so.  Figure~\ref{fg:evalues} shows a typical eigenspectrum for the system in Eqs.~\eqref{Fotsos1}-\eqref{Fotsos2} on two subdomains.  The eigenvalues with positive real parts clearly indicate instability.

\subsection{Second-Order Penalty Method} \label{MainResult}

\noindent We will now derive a way to impose penalty boundary conditions in the second-order system that yields a robust, stable result.  For the semi-discrete problem, we once again choose Gauss-Legendre-Lobatto collocation points.  We begin by writing the second-order equations in the form
\begin{flalign}
\dot \psi_i =&\, -\pi_i + p, \label{UndetEq1}\\
\dot \pi_i =&\, -\psi_i^{\prime\prime} + q, \label{UndetEq2}
\end{flalign}
where $p$ and $q$ represent as yet undetermined penalties.  The semi-discrete energy is
\beq
E = \frac{1}{2}\left[\langle \pi,\pi \rangle + \langle \psi^\prime,\psi^\prime \rangle\right]. \label{1dEnergy}
\eeq
Taking the time derivative, we find
\beq
\dot E = - \psi_i^\prime \pi_i \big|_0^N + \psi_i^\prime\, p_i \big|_0^N + \langle \pi,q \rangle - \langle \psi^{\prime\prime},p \rangle, \label{Eeq1}
\eeq
where we have used summation by parts in the first two terms.  The first term is like the continuum result:
\beq
- \psi_i^\prime \pi_i \big|_0^N = \frac{1}{4} \sum\limits_{i=0,N} (U_-^2 - U_+^2). \label{ContPart}
\eeq
Since the projection of boundary conditions from characteristic to fundamental variables is unambiguous in the variable $\pi$, we will write the penalty $q$ as in Eq.~\eqref{Fotsos2}:
\beq
q = \frac{a}{\omega}\,(\delta_{i0}+\delta_{iN})\, \delta U_-, 
\eeq
where $a$ is an undetermined constant and $\omega$ is the quadrature weight at $x=\pm1$.  The factor $1/\omega$ is explicitly written in anticipation of its cancellation when evaluating the third term in Eq.~\eqref{Eeq1}:
\beq
\langle \pi, q \rangle \, =\, a \pi_0 \,\delta U_-^0 + a \pi_N \delta U_-^N. 
\eeq
If we also choose the penalty $p$ in Eq.~\eqref{UndetEq1} to have a similar value on the boundary
\beq
p = a \,\delta U_-, \label{pBdry}
\eeq
then the second and third terms in Eq.~\eqref{Eeq1} combine to form the expression
\beq
\psi_i^\prime\, p_i \big|_0^N + \langle \pi,q \rangle = \frac{a}{2} \sum\limits_{i=0,N} (U_-^\text{BC}{}^2 - U_-^2 - \delta U_-^2).
\eeq
Note that we do not define $p$ on the boundary with a factor of $1/\omega$, because the second term in Eq.~\eqref{Eeq1} arises out of summation by parts as opposed to being picked out from the discrete sum by a Kronecker delta.

The stumbling block in this energy analysis is the last term in Eq.~\eqref{Eeq1}, whose appearance is inevitable because of the derivatives in the definition of energy in Eq.~\eqref{1dEnergy}.  Such a term did not arise in the first-order energy estimate, precisely because the first-order energy in Eq.~\eqref{FoshSDE} did not contain any derivatives.  Fortunately, it turns out we can eliminate the inner product $\langle \psi^{\prime\prime},p \rangle$ by allowing the penalty $p$ to be non-zero throughout the domain and by constructing it to be orthogonal to $\psi^{\prime\prime}$.  

The scalar field $\psi$ in the semi-discrete solution is an interpolating $N^\text{th}$-order polynomial~\cite{Boyd2001}.  Therefore, $\psi^{\prime\prime}$ is an $N\!-\!2$ order polynomial, and the product $\psi^{\prime\prime}p$ is at most a polynomial of order $2N\!-\!2$.  It follows that the quadrature integral is exact:
\beq
\langle \psi^{\prime \prime},p \rangle = \int\limits_{-1}^{1} \psi^{\prime \prime}(x)p(x)\,dx.
\eeq
This inner product will automatically vanish if the penalty $p$ is a linear combination of the Legendre polynomials $P_N(x)$ and $P_{N-1}(x)$, which are orthogonal to any polynomial of degree $N\!-\!2$ or less.  We are therefore provided with two degrees of freedom for constructing the function $p$, which is sufficient to satisfy the boundary values defined in Eq.~\eqref{pBdry}.  We make use of the following polynomials, constructed to take the values $0,1$ at $x=\pm1$:
\begin{flalign}
f(x) =&\, \frac{1}{2}(-1)^N \left[ P_N(x)- P_{N-1}(x) \right], \label{fdef}\\
g(x) =&\, \frac{1}{2} \left[P_N(x) + P_{N-1}(x) \right]. \label{gdef}
\end{flalign}
If we now define the penalty $p$ to be
\beq
p = p_0\,f(x) + p_N\,g(x), \label{pdef}
\eeq
where $p_0$ and $p_N$ represent the endpoint values of Eq.~\eqref{pBdry}, then the penalty function $p$ will have the correct boundary values while also satisfying
\beq
\langle \psi^{\prime\prime}, p \rangle = 0.
\eeq
Putting things together, we now obtain
\beq
\dot E =\frac{1}{4}\sum\limits_{i=0,N} \big[(1-2a )U_-^2  -U_+^2 + 2a(U_-^\text{BC}{}^2 - \delta U_-^2) \big], \label{1dedota}
\eeq
which is just like Eq.~\eqref{FoshEdot} for the first-order system with $2a \leftrightarrow c\,\omega$.  The conclusions reached previously for $c$ therefore carry over: a multi-domain problem with arbitrary outer boundary conditions is stable only if $a=1/2$.  The second-order system with penalties is thus
\begin{flalign}
\dot \psi_i =&\, -\pi_i - \frac{1}{2}\left[f(x)\,\delta U_-^0 + g(x)\,\delta U_-^N\right], \label{FotsosG1} \\ \noalign{\vskip 3pt}
\dot \pi_i =&\, -\psi_i^{\prime\prime} + \frac{1}{2\omega}\left[\delta_{i0}\,\delta U_-^0 
+ \delta_{iN}\,\delta U_-^N\right]. \label{FotsosG2}
\end{flalign}
Of course, one needs to be concerned not only with \textit{stability}, but also \textit{consistency}---that is, the system should reproduce the continuum equations in the limit as $N\rightarrow \infty$.  The penalty $p$ on the $\dot \psi$ equation in Eq.~\eqref{FotsosG1} is applied throughout the domain and not only on the boundaries.  Moreover, it does not scale as $N^2$, so consistency might seem dubious.  However, the penalty on $\dot \pi$ in Eq.~\eqref{FotsosG2} does scale as $N^2$ and is applied only on the boundaries.  Therefore, the condition $\delta U_- \rightarrow 0$ on the boundary as $N\rightarrow \infty$ is enforced.  This also implies $p \rightarrow 0$ in turn, so consistency follows.  
	
\begin{figure}[t!]
\centering
\includegraphics[width=\columnwidth]{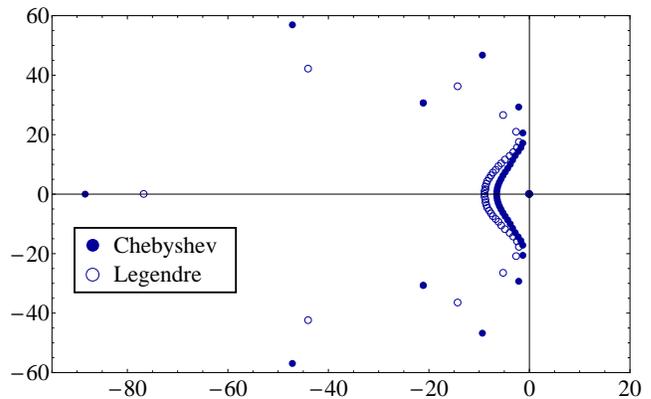}
\caption{Eigenvalues in the complex plane of the second-order system in Eqs.~\eqref{FotsosG1}-\eqref{FotsosG2} on two subdomains covering $[-1,1]$, with an inner boundary at $x=0$, outer boundary conditions $U_-^\text{BC}=0$, and $N+1=11$ grid points per subdomain.  Results for Legendre- and Chebyshev-Lobatto grids are shown for comparison.}
\label{fg:evalues2}
\end{figure}

Although the second-order energy argument was performed on Legendre points, the equations can be implemented on any grid, just as in the first-order system discussed in Section~\ref{Fosh1D}.  Eigenvalues of the fully discrete system imply stability here as well, as shown in Fig.~\ref{fg:evalues2} for a representative two-domain problem.  The spectral radius is somewhat larger than in the first-order spectrum in Fig.~\ref{fg:evalues}.  However, we find that differences in CFL conditions essentially disappear for more general systems, including the three-dimensional wave equation in flat or curved space.

%======================== FIGURE Fotsos Good Eigenvalues ===============================
%============================================================================

\section{Three-Dimensional Wave Equation} \label{3DSWF}

\noindent Let us now consider the three-dimensional second order in space wave equation 
\begin{flalign}
\dot \psi =&\, -\pi, \\
\dot \pi =&\, -\d_i\d^i\psi.
\end{flalign}
The characteristic modes and speeds of this system are
\begin{flalign}
\qquad \qquad U_\psi =&\, \psi, & \lambda &= 0, \qquad \qquad \label{Upsidef}\\
\qquad \qquad U_\pm  =&\, \pi \pm n^i\d_i\psi, & \lambda &= \pm 1, \qquad \qquad \label{Upmdef} \\
\qquad \qquad U^0_i  =&\, \d_i\psi - n_i n^j\d_j\psi, & \lambda &= 0, \qquad \qquad \label{U0def}
\end{flalign}
where $n^i$ is the outward-directed unit normal to the boundary.  These are the same as the characteristic variables of the first-order system obtained by defining $\phi_i \equiv \d_i \psi$. Usually, one thinks of characteristic variables as being defined only for first-order systems, but they can be generalized to second-order systems.  One way to do this is to define the second-order modes as those combinations $U$ of variables $(\pi, \d_i\psi)$ that satisfy
\beq
\dot U = -\lambda \,n^i \d_i U + \ldots,  \label{FotsosCharVar}
\eeq
where the dots represent derivatives transverse to $n^i$ plus lower order terms~\cite{GundlachMartinGarcia2006}. As a consequence of this definition, the transverse derivatives $U^0_i$ are automatically zero-speed modes (in fact, they can be given arbitrary speeds).  Moreover, the characteristic variables in Eq.~\eqref{Upmdef} are unique only up to addition of these zero-speed modes.  For example, we could redefine $U_\pm$ as $U_\pm + X^i U^0_i$ for arbitrary (fixed) $X^i$.  As discussed in Ref.~\cite{GundlachMartinGarcia2006}, this ambiguity is removed for a symmetric hyperbolic system by requiring the existence of a conserved energy that is quadratic in the modes.  Here, that amounts to taking the definitions in Eqs.~\eqref{Upsidef}-\eqref{U0def} as they are.  The conserved energy density is
\beq
\epsilon = \frac{1}{2}(\pi^2 + \d^i\psi\,\d_i\psi). \label{3dSWFenergy}
\eeq
Note that this energy is indeed quadratic in terms of the characteristic modes:
\beq
\epsilon = \frac{1}{2}\big(U_+^2 + U_-^2\big) + U^{0i}U^0_i.
\eeq
In analogy with the one-dimensional case in Eq.~\eqref{1dflux}, the flux is
\beq
\dot E = \frac{1}{4} \int_{\d \Omega} (U_-^2 - U_+^2)\,d^2x,
\eeq
where $\d \Omega$ represents the boundary of the domain. 

Now consider the semi-discrete problem in three-dimensions.  We encounter a few issues in generalizing from the one-dimensional case.  For one, if the boundary of the domain contains edges or corners, the normal vectors there (and hence characteristic modes) are not well-defined.  For reasons that will become clear below, we resolve this ambiguity by defining the normal vectors as follows.  We will use upper case ${\bf N}$ and lower case ${\bf n}$ to denote the unnormalized and unit normal vectors, respectively.  For simplicity, suppose the domain $\Omega$ is a cube with $x,y,z \in [-1,1]$.  On boundary faces (codimension $1$), one coordinate is fixed (e.g.\ the $x=+1$ face).  We define face normals on a boundary with a fixed $i^{\text{th}}$ coordinate as 
\beq
{\bf N} = \omega_j \,\omega_k\, {\bf n},
\eeq
where $\omega_j$ and $\omega_k$ are the quadrature weights (see Appendix~\ref{appGLLnotes}) corresponding to the two free dimensions, and ${\bf n}$ is the usual (Cartesian) unit normal vector in the $i^\text{th}$ direction.   On edges and corners, the normal vector is defined to be the sum of the normals to the adjacent boundary faces.  For example, the normal vector at the corner $(x,y,z) = (1,1,1)$ is defined to be
\beq
{\bf N} = \omega_y\, \omega_z\, {\bf \hat x} + \omega_x\, \omega_z \,{\bf \hat y} + \omega_x\, \omega_y \,{\bf \hat z}. \label{normal}
\eeq
The second-order system with penalty functions $p,q$ is
\begin{flalign}
\dot \psi =&\, -\pi + p, \label{3deq1}\\
\dot \pi =&\, -\d_i\d^i\psi + q , \label{3deq2}
\end{flalign}
where for conciseness we have suppressed indices representing grid values (e.g.\ $\psi = \psi_{ijk}$).  Consider the semi-discrete energy
\beq
E = \frac{1}{2}\left[ \langle \pi,\pi \rangle + \langle \d^i\psi, \d_i\psi \rangle \right]. \label{3dSdE}
\eeq
The computation of $\dot E$ proceeds analogously to the one-dimensional case, except for complications due to the corners and edges.  To see this, consider the following term that arises in taking the time derivative of Eq.~\eqref{3dSdE}:
\beq
\langle \d^i \psi, \d_i \pi \rangle \equiv \sum\limits_{i,j,k} \omega_i\omega_j\omega_k \, \d^l\psi\, \d_l \pi. \label{3dEdotpart}
\eeq
We use summation by parts in this expression and obtain three boundary terms---one for each $l$.  For example, from $l=z$ we get 
\beq
\langle \d^z\psi,\d_z\pi\rangle = \sum\limits_{i,j} \omega_i\omega_j\,\big[\d_z\psi\, \pi\big]_{z=-1}^{+1} - \langle \d^z\d_z\psi,\pi\rangle.
\eeq
Each point on an edge receives a contribution from two such boundary terms, while points on corners get a contribution from all three.  On the cube at the corner point $(1,1,1)$, for example, the value obtained is
\beq
\big(\omega_x\omega_y\,\d_z\psi + \omega_x\omega_z\,\d_y\psi + \omega_y\omega_z\,\d_x\psi\big)\,\pi. \label{cornerterm}
\eeq
We would like to be able to write this in terms of characteristic modes as
\beq
N^i\d_i\psi\,\pi = \frac{|{\bf N}|}{4}\big(U_+^2 - U_-^2),
\eeq
and this is precisely the reason for the definition of normal vectors on edges and corners given above.  Thus, Eq.~\eqref{3dEdotpart} can be written
\beq
\langle \d^i \psi, \d_i \pi \rangle = \frac{1}{4} \sum\limits_{\d \Omega} |{\bf N}| \big( U_+^2 - U_-^2 \big) - \langle \d^i\d_i \psi, \pi \rangle, \label{3dcontpart}
\eeq
where the sum is over all boundary points, including edges and corners.  The magnitude of the normal vector $|{\bf N}|$ encodes the appropriate quadrature weight factors for boundaries of any codimension.  

In a similar way, the terms in $\dot E$ due to the penalty $p$ in Eq.~\eqref{3deq1} can be written
\beq
\langle \d^i \psi, \d_i p \rangle = \sum\limits_{\d \Omega} |{\bf N}| n^i\d_i \psi\, p - \langle \d^i\d_i \psi, p \rangle.
\eeq
The penalty $q$ in Eq.~\eqref{3deq2} is applied only on the boundary, where it takes the value
\beq
q\Big|_{\d \Omega} = \frac{1}{2}\frac{|{\bf N}|}{\omega_x \omega_y \omega_z}\delta U_-.
\eeq
On a boundary face with fixed $i^\text{th}$-coordinate, this reduces to
\beq
q = \frac{1}{2}\frac{1}{\omega_i}\delta U_-,
\eeq
just as in the one-dimensional system.  Assuming the boundary values of $p$ satisfy
\beq
p\Big|_{\d \Omega}= -\frac{1}{2}\delta U_-, \label{3dpbdry}
\eeq
the penalty contributions to $\dot E$ combine to give
\beq
\dot E_\text{penalties} = \frac{1}{2}\sum\limits_{\d \Omega} | {\bf N} | U_- \delta U_- - \langle \d^i\d_i \psi, p \rangle.
\eeq
With everything included, the energy flux is
\beq
\dot E = \frac{1}{4} \sum\limits_{\d \Omega} |{\bf N}| \left( U_-^{\text{BC}}{}^2 - U_+^2 - \delta U_-^2 \right) - \langle \d^i \d_i \psi, p \rangle.  \label{3dedot}
\eeq
We would like to eliminate the last term with an appropriate choice of bulk penalty function $p$, as was done in the one-dimensional case.  The most obvious generalization of the one-dimensional approach would be to construct $p$ out of polynomials $\theta_n$ satisfying $\langle \d^i\d_i\psi, \theta_n \rangle = 0 $.  Unfortunately, this cannot be done.  There are in general only about $2N^2$ such functions $\theta_n$---not enough to satisfy $6N^2\!+\!2$ boundary conditions (a proof is provided in Appendix~\ref{appDOFProof}).  

Alternatively, one could seek a solution by allowing the penalty $p$ to depend explicitly on the scalar field $\psi$.  One way of doing this is to split the offending inner product term of Eq.~\eqref{3dedot} into contributions from the boundary and the interior of the domain:
\beq
\langle \Delta \psi, p \rangle = \langle \Delta \psi, p \rangle \big|_{\d \Omega} + \langle \Delta \psi, p \rangle \big|_{\text{interior}}, \label{innerprodsplit}
\eeq
where $\Delta \psi \equiv \d^i \d_i \psi$.  Considering the values of $p$ on the boundary to be specified by Eq.~\eqref{3dpbdry}, the first term on the right-hand side of Eq.~\eqref{innerprodsplit} is fixed.  We can then define $p$ on the interior of the domain to be 
\beq
p_\text{interior} \equiv -\frac{\langle \Delta \psi, p \rangle \big|_{\d \Omega}}{\langle \Delta \psi, \Delta \psi \rangle \big|_\text{interior}}\,\Delta \psi, \label{normp}
\eeq
provided $\Delta \psi_\text{interior} \ne 0$.  With this definition, the discrete sum over the interior cancels the sum over the boundary in Eq.~\eqref{innerprodsplit}, and the inner product $\langle \Delta \psi, p \rangle$ vanishes.  If $\Delta \psi \equiv 0$, we cannot use Eq.~\eqref{normp}, but in this case there would be no need since then $\langle \Delta \psi, p \rangle \equiv 0$.  One way this recipe could fail is if $\Delta \psi$ vanishes on the interior of the domain but not on the boundary, although we believe this to be very unlikely in an actual numerical simulation.  We find that while this method yields stability, it suffers from lack of convergence as resolution is increased.  However, as we have not experimented extensively with approaches like this, further investigation may be worthwhile.

Abandoning any explicit dependence on $\psi$ in the penalties, we seek instead to construct $p$ so as to \textit{minimize} the inner product $\langle \d^i \d_i \psi, p \rangle$ of Eq.~\eqref{3dedot}.  It turns out this can be done by using a penalty constructed out of the same functions $f,g$ defined in Eqs.~\eqref{fdef}-\eqref{gdef} for the one-dimensional problem.  Here we will give a summary of the result; a derivation is provided in Appendix~\ref{appBulkPenalty}.  We define one-dimensional functions $f,g$ along each dimension and write their grid values as $f_i=f(x_i)$, $f_j=f(y_j)$, $f_k=f(z_k)$ (and similarly for $g$).  Assuming the values of the penalty function $p$ on the domain boundary $\d \Omega$ are given, the grid values on the interior of the domain are
\beq
\begin{split}
p_{ijk} =&\> p_{0jk}\,f_i + p_{N\!jk}\, g_i + p_{i0k}\,f_j + \ldots \qquad \text{(faces)} \label{bulkp} \\ 
&- p_{00k}\,f_i f_j - p_{N0k} \,g_i f_j - \ldots \qquad \>\>\>\, \text{(edges)}  \\ 
&+ p_{000}\,f_i f_j f_k + p_{N\!00}\, g_i f_j f_k + \ldots \,\, \text{(corners)} 
\end{split}
\eeq
The bulk penalty picks up a contribution from each boundary face, edge, and corner.  The assumption of a cubic domain is not a limitation, as it is straightforward to generalize this procedure to other domains.  With this choice of penalty, $\dot E$ is again given by Eq.~\eqref{3dedot}, and the last term in Eq.~\eqref{3dedot} vanishes in the limit $N\rightarrow\infty$ (see the discussion at the end of Appendix~\ref{appBulkPenalty}).

Therefore, while not strictly stable, the system is \textit{asymptotically} stable.  Collecting results, we conclude that the second-order system is
\begin{flalign}
\dot \psi =&\, -\pi - \frac{1}{2}\delta U_-, \label{3deq3}\\
\dot \pi =&\, -\d_i\d^i\psi + \frac{1}{2}\frac{|{\bf N}|}{\omega_x \omega_y \omega_z}\delta U_-, \label{3deq4}
\end{flalign}
where the penalties represent boundary values, the penalty on the first equation is applied throughout the interior of the domain via Eq.~\eqref{bulkp}, and the normal vector ${\bf N}$ is defined as in Eq.~\eqref{normal}.

\subsection*{Numerical Tests}

\begin{figure}[t!]
\centering
\includegraphics[width=\columnwidth]{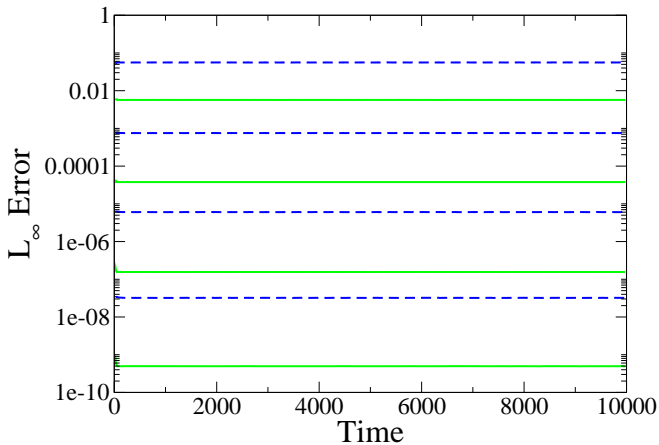}
\caption{$L_\infty$ error of a plane wave $\psi({\bf x},t) = \sin ({\bf k}\! \cdot \!{\bf x} - \omega\,t)$ evolved using the second-order system (solid green).  Results of equivalent first-order evolutions are plotted for comparison (dashed blue).  The domain consists of $27$ identical cubic subdomains covering the region $x,y,z \in [-15,15]$, and the successive resolutions have $5,7,9$, and $11$ Legendre-Lobatto grid points per subdomain along each dimension. In this test, ${\bf k} = (.3,.2,.1)$ and $\omega=|{\bf k}|$.  The $L_\infty$ error is a moving average over an interval $\Delta t = 50$, which includes $50$ data points.}
\label{fg:FlatConv}
\end{figure}

\noindent The three-dimensional wave equation in Eqs.~\eqref{3deq3}-\eqref{3deq4} with bulk penalty given by Eq.~\eqref{bulkp} is found to be robust, stable, and convergent in all of our tests.  We have run simulations on multiple spherical shell, cylindrical shell, and cubic subdomains.  As an example, Fig.~\ref{fg:FlatConv} shows the $L_\infty$ error (maximum nodal error) of a sinusoidal plane wave propagating through a domain consisting of $27$ cubic subdomains.  

In this example, only the boundary-face part of the bulk penalty in Eq.~\eqref{bulkp} is used, and normal vectors are defined as in Eq.~\eqref{normal}.  Empirically, we find that the bulk penalties associated with edges and corners are not needed in this example.  The incoming mode at an interface boundary is supplied by the adjacent subdomain, while at outer boundaries it is computed from the analytical solution.  Time-stepping is performed using an explicit fourth-order Runge-Kutta method.  The results of an equivalent first-order evolution are plotted for comparison.

%====================== FIGURE 3D flat space convergence ===============================
%===========================================================================

A few empirical observations are worth noting.  In practice, we find that the bulk penalty terms arising from edges and corners in Eq.~\eqref{bulkp} are not actually necessary for obtaining stable, convergent evolutions.  In all the tests we have performed for scalar waves in flat space, the terms due to the faces of the boundary are sufficient.  However, the additional terms in Eq.~\eqref{bulkp} may need to be included for complicated domain decompositions or in curved space applications.

For more general systems of quasi-linear wave equations (such as Einstein's equations in generalized harmonic form~\cite{Lindblom2006}), we find that it is sometimes necessary to include a boundary term enforcing continuity of the field $\psi$ in the penalty.  That is, one makes the replacement $\delta U_- \rightarrow \delta U_- + \delta \psi$ in the penalties.  In the tests that we have performed, this is not required for a simple wave equation in flat (or curved) space.

An alternative to defining unique normal vectors on corners and edges is to use a so-called multi-penalty method.  With a multi-penalty method, boundary conditions (and hence penalties) on edges and corners are defined to be the sum of those from the adjacent boundary faces.  While this has the advantage of avoiding some of the issues with corners and edges, it makes obtaining analytical results such as Eq.~\eqref{3dedot} more difficult.  Although we have not yet fully tested this alternative in curved space applications, we find that the multi-penalty method performs equally well for scalar waves in flat space.

\section{Wave Equation on Curved Background} \label{3DSWC}

\noindent In this section we consider the application of the new penalty method to the evolution of a scalar wave on a fixed, curved background spacetime:
\beq
\nabla_\mu\nabla^\mu \psi = 0, \label{CurvedWaveEq}
\eeq 
where $\nabla_\mu$ is the four-dimensional covariant derivative.  In rewriting this equation as a first-order system, we use the standard $3+1$ splitting of the metric:
\beq
ds^2 = -\alpha^2 dt^2 + \gamma_{ij}(dx^i + \beta^i dt)(dx^j+\beta^j dt),\label{3p1metric}
\eeq
where $\alpha$ is the lapse function, $\beta^i$ is the shift, and $\gamma_{ij}$ is the three-dimensional metric intrinsic to the constant time spatial hypersurfaces.  It is assumed that $\alpha >0$ and that the three-metric $\gamma_{ij}$ is positive definite.

The wave equation in Eq.~\eqref{CurvedWaveEq} can be rewritten in a standard way~\cite{Scheel2004} as the first-order system
\begin{flalign}
\dot \psi =&\, -\alpha\pi + \beta^i\d_i\psi, \label{FoshCurvedEq1}\\
\dot \pi =&\, -\alpha \gamma^{ij}\d_i\phi_j + \beta^i\d_i\pi + \alpha K \pi + \alpha J^i \phi_i, \label{FoshCurvedEq2} \\
\dot \phi_i =&\,- \alpha \d_i \pi - \pi\d_i \alpha + \phi_k\d_i\beta^k + \beta^k\d_k \phi_i. \label{FoshCurvedEq3}
\end{flalign}
Equation~\eqref{FoshCurvedEq1} is just the definition of the variable $\pi$. As usual, the spatial derivative variable is defined as
\beq
\phi_i \equiv \d_i \psi.
\eeq
The quantities $K$ and $J^i$ in Eq.~\eqref{FoshCurvedEq2} are purely functions of the background spacetime:
\begin{flalign}
K \equiv&\, - \frac{1}{\alpha \gamma^{1/2}} \Big[\d_0 \gamma^{1/2} - \d_i\big(\gamma^{1/2}\beta^i\big)\Big], \\
J^i \equiv&\, - \frac{1}{\alpha \gamma^{1/2}} \,\d_j \big(\alpha\gamma^{1/2}\gamma^{ij}\big), 
\end{flalign}
where $\gamma \equiv \det \gamma_{ij}.$  In deriving Eq.~\eqref{FoshCurvedEq3}, the equivalence of interchanging indices in
\beq
\d_i\phi_j = \d_j \phi_i
\eeq
has been assumed.  This reduction to first order has therefore introduced two constraints to the system: $\mathcal{C}_i = \mathcal{C}_{ij} = 0$, where
\begin{flalign}
\mathcal{C}_i & \equiv \phi_i - \d_i\psi, \label{Con1} \\
\mathcal{C}_{ij} & \equiv \d_i\phi_j - \d_j\phi_i.\label{Con2}
\end{flalign}
The second-order in space equations are
\begin{flalign}
\dot \psi =&\, -\alpha\pi + \beta^i\d_i\psi, \label{CurvedEq1}\\
\dot \pi =&\, -\alpha \gamma^{ij}\d_i\d_j\psi + \beta^i\d_i\pi + \alpha K \pi + \alpha J^i \d_i\psi. \label{CurvedEq2}
\end{flalign}
This system avoids the introduction of the constraints in Eqs.~\eqref{Con1}-\eqref{Con2} as well as the third set of evolution equations in Eq.~\eqref{FoshCurvedEq3}. 

The characteristic variables and speeds of the second-order system are the same as those of the equivalent first-order reduction with $\phi_i \rightarrow \d_i\psi$:
\begin{flalign}
\qquad U_\psi =&\, \psi, & \lambda_0 &= -n^k\beta_k \label{UpsidefSWC} \\
\qquad U_\pm  =&\, \pi \pm n^i\d_i\psi, & \lambda_\pm\! &= \pm \alpha - n^k\beta_k, \label{UpmdefSWC}\\
\qquad U^0_i  =&\, \d_i\psi - n_i n^j\d_j\psi, & \lambda_0 &= -n^k\beta_k,\label{U0defSWC} 
\end{flalign}
where $n^i$ is the outward-directed unit normal vector to the boundary of the three-dimensional spatial domain.  These are the same as the characteristic modes of the scalar wave in flat space in Eqs.~\eqref{Upsidef}-\eqref{U0def}, with modified characteristic speeds.  As discussed in Section~\ref{3DSWF} above, the ``zero-speed'' modes $U^0_i$ can be considered to have arbitrary speeds in the second-order system~\cite{GundlachMartinGarcia2006}.  The speeds $\lambda_0$ given above are chosen to be the same as those of the corresponding first-order system.  Additionally, these are the coefficients that appear in the boundary flux of the energy, and in this sense they are the preferred choice.

\subsection*{Continuum Energy Estimate}

\noindent The energy density for this system is the same as for the flat space scalar wave in Eq.\eqref{3dSWFenergy}:
\beq
\epsilon = \frac{1}{2} (\pi^2 + \d^i\psi\d_i\psi). \label{CurvedEps}
\eeq
The energy flux is found by computing the time derivative of the energy
\beq
E = \int_\Omega \epsilon\, \gamma^{1/2} d^3x,
\eeq
where $\Omega$ is the spatial domain under consideration and $\gamma^{1/2}d^3x$ is the volume element.  In addition to a boundary flux, differentiating the energy gives rise to volume terms that depend on various derivatives of the background ($\d_i\alpha,\, \d_i\beta_j$, or $\d_i\gamma_{jk}$).  However, these volume terms can all be bounded by multiples of the energy itself (or neglected entirely in the constant-coefficient approximation), which is all that is required for proving well-posedness.  One therefore obtains
\beq
\dot E \le - \frac{1}{4}\int_{\d\Omega} F^n\, \sigma^{1/2} d^2x \,+\, k\,E, \label{EdotSWC}
\eeq
for some constant $k \ge 0$.  The flux integrand is
\beq
F^n = \lambda_- U_-^2 + \lambda_+ U_+^2 + 2 \lambda_0 \,U^{0j}U^0_j, \label{FluxSWC}
\eeq
and the element of area in Eq.~\eqref{EdotSWC} is $\sigma^{1/2}d^2x$, where $\sigma \equiv \det \sigma_{ij}$ and $\sigma_{ij}$ is the intrinsic metric on the boundary surface.  The continuum problem is therefore well-posed with boundary conditions that control incoming modes (those with $\lambda < 0$).  For a timelike boundary, a boundary condition is needed on $U_-$ and possibly on $U^0_i$, depending on the sign of $\lambda_0$.  For a spacelike boundary, either all modes are incoming, or all modes are outgoing and no boundary conditions are required (e.g.\ on an excision boundary inside the horizon of a black hole).  

We could also have included a term $a^2 \psi^2$ in the energy density, replacing Eq.~\eqref{CurvedEps} by
\beq
\epsilon = \frac{1}{2} ( a^2\, \psi^2 + \pi^2 + \d^i\psi\d_i\psi). \label{CurvedEpsPsi2}
\eeq
This would give an additional term in $\dot E$:
\beq
\int_\Omega a^2 \psi \dot \psi \,\gamma^{1/2} d^3x = \int_\Omega a^2 \psi \big(\beta^i\d_i\psi-\alpha \pi \big)\gamma^{1/2} d^3x.\label{SWCpsi2Epart} 
\eeq
Integrating by parts in the first term on the right-hand side yields
\beq
\frac{a^2}{2} \int_{\d \Omega} n_i\beta^i\, \psi^2 \,\sigma^{1/2} d^2x 
- \frac{a^2}{2} \int_\Omega \psi^2 \,\d_i \big(\beta^i \gamma^{1/2}\big) d^3x. \label{psi2Bdryflux} 
\eeq
The latter term in this expression can be bounded by a multiple of the energy, while the first term contributes to the boundary flux.  It may seem, then, that including the term $a^2 \psi^2$ in the energy density would require the flux $F^n$ of Eq.~\eqref{FluxSWC} to be modified.  However, the entire right-hand side of Eq.~\eqref{SWCpsi2Epart} can in fact be bounded in the volume.  Making use of the relation $(\hat \beta^i \d_i \psi)^2 \le \d^i\psi\d_i\psi$, we find
\beq
\begin{split}
\int_\Omega a^2 \psi \big(\beta^i\d_i\psi - & \alpha \pi \big)\gamma^{1/2} d^3x \\
& \quad \le a\, (\alpha_\text{max}+|\beta|_\text{max}) E. \label{psi2Inequality}
\end{split}
\eeq
The addition of a term $a^2\psi^2$ to the energy density therefore requires the constant $k$ in Eq.~\eqref{EdotSWC} to be modified, but not the flux $F^n$.  Consequently, our conclusions about well-posedness and boundary conditions remain unchanged.  It is interesting to note, however, that the same does not hold for the first-order system of Eqs.~\eqref{FoshCurvedEq1}-\eqref{FoshCurvedEq3}, because the first-order energy corresponding to Eq.~\eqref{CurvedEpsPsi2} controls $\phi_i$, but not $\d_i \psi$ (and therefore the inequality in Eq.~\eqref{psi2Inequality} does not follow).

\subsection*{Semi-discrete Energy Estimate}

\noindent The penalties in the semi-discrete equations need to be slightly modified from those of the flat space scalar wave system in Eqs.~\eqref{3deq3}-\eqref{3deq4}.  To see how, consider the semi-discrete equations corresponding to Eqs.~\eqref{CurvedEq1}-\eqref{CurvedEq2} with penalty functions $p, q$:
\begin{flalign}
\dot \psi_i =&\, -\alpha\pi_i + \ldots + p, \label{SDCurvedEq1}\\
\dot \pi_i =&\, -\alpha \gamma^{jk}\d_j\d_k\psi_i + \ldots + q. \label{SDCurvedEq2}
\end{flalign}
As usual, is it to be understood that the fields represent grid values (e.g.\ $\psi = \psi_{ijk}$), and differentiation is implemented, for example, by matrix multiplication.  For simplicity, we will assume that the physical domain under consideration has been mapped to the cube $\Omega$ with $x,y,z \in [-1,1]$.  We will also assume for now that the boundary is timelike, with $U_-$ the only incoming mode.  As in the flat space scalar wave system, the penalties will thus be proportional to $\delta U_- \equiv U_-^\text{BC} - U_-$.  The semi-discrete energy is
\beq
E = \frac{1}{2} \big[ \langle \pi,\pi \rangle + \langle \d^i\psi,\d_i\psi \rangle \big], \label{SDEnergy}
\eeq
where the discrete inner product is now defined by, for example
\beq
\begin{split}
\langle \pi,\pi \rangle &\equiv \sum\limits_{ijk} \omega_i\omega_j\omega_k \, \pi^2 \gamma^{1/2} \\
&\simeq \int_\Omega \pi^2\,\gamma^{1/2} d^3x.
\end{split}
\eeq
Because of the presence of $\gamma^{1/2}$ in the volume element, the quadrature integrals we encounter will in general no longer be exactly equal to the continuum integrals.  

The time derivative of the semi-discrete energy in Eq.~\eqref{SDEnergy} separates as usual into a continuum-like part plus a contribution from the penalties:
\beq
\dot E = \dot E_\text{continuum} + \dot E_\text{penalties},
\eeq
where the penalty contribution is
\beq
\dot E_\text{penalties} = \langle \pi, q \rangle + \langle \d^i\psi,\d_i p \rangle. \label{EdotPenalties}
\eeq
Assuming the penalty $q$ is defined as in Eq.~\eqref{3deq4} except for an overall factor $q_0$, we have
\beq
\langle \pi,q\rangle = \frac{1}{2}\sum\limits_{\d\Omega} |{\bf N}| q_0\,\pi\,\delta U_-\,\gamma^{1/2}, \label{qEdotSWC}
\eeq
where $|{\bf N}|$ is the magnitude (now with respect to $\gamma_{ij}$) of the normal vector defined as in Eq.~\eqref{normal}.  The second term in Eq.~\eqref{EdotPenalties} gives
\beq
\langle \d^i\psi ,\d_i p \rangle = \sum\limits_{\d \Omega} |{\bf N}| n^i \d_i \psi \,p \,\gamma^{1/2} - \langle \nabla_j\nabla^j\psi,p \rangle, \label{pEdotSWC}
\eeq
where $n^i$ is the unit normal vector to the boundary, and $\nabla_j$ is the three-dimensional covariant derivative associated with $\gamma_{ij}$.  With the penalty function $p$ constructed in the volume according to Eq.~\eqref{bulkp}, the last term in Eq.~\eqref{pEdotSWC} asymptotically vanishes as in the flat-space case, and we will therefore neglect it.  Assuming that $p$ has the same value as in Eq.~\eqref{3deq3} apart from an overall factor $p_0$, it follows that
\beq
\langle \d^i\psi ,\d_i p \rangle = -\frac{1}{2} \sum\limits_{\d\Omega} |{\bf N}|\, n^i \d_i \psi \,\,p_0\,\delta U_- \,\gamma^{1/2}. 
\eeq
If we choose $q_0 = p_0$, then Eq.~\eqref{EdotPenalties} for the penalty contribution to $\dot E$ becomes
\beq
\dot E_\text{penalties} = \frac{1}{2}\sum\limits_{\d\Omega} |{\bf N}| p_0 \,U_- \delta U_-\, \gamma^{1/2}.
\eeq
Setting $p_0 = |\lambda_-|$, we obtain the semi-discrete energy estimate
\beq
\dot E \le -\frac{1}{4}\sum\limits_{\d\Omega} |{\bf N}|\, F^n\, \gamma^{1/2} \,+\, k\,E, \label{SDEdotSWC}
\eeq
for some constant $k\ge0$.  The flux integrand is
\beq
F^n = \lambda_- \big(U_-^\text{BC}{}^2 - \delta U_-^2 \big) + \lambda_+ U_+^2 + 2 \lambda_0 U^{0i}U^0_i, \label{FluxSWCsd}
\eeq
which resembles the continuum result of Eq.~\eqref{FluxSWC} with the addition of the negative term proportional to the mismatch of characteristic modes $\delta U_-^2$.  The sum over the boundary can be rewritten as
\beq
\sum\limits_{\d\Omega} |{\bf N}| \gamma^{1/2}\, (\,\cdot\, ) = \sum\limits_{\d\Omega} |{\bf \tilde N}|_\text{E} \sigma^{1/2} \,(\,\cdot\,), 
\eeq
where $|{\bf \tilde N}|_\text{E}$ is the magnitude (with respect to a Euclidean metric) of the normal one-form corresponding to ${\bf N}$, and $(\,\cdot\,)$ represents any integrand.  In this form, the similarity to the surface integral 
\beq
\int_{\d \Omega} (\,\cdot\,) \,\sigma^{1/2} d^2x
\eeq
in the continuum result of Eq.~\eqref{EdotSWC} is evident.

We have assumed that $U_-$ is the only incoming mode, and in this case Eq.~\eqref{SDEdotSWC} shows that the semi-discrete system is asymptotically well-posed.  If $\lambda_0 < 0$, then the boundary flux in Eq.~\eqref{FluxSWCsd} implies that a boundary condition is required to control $U^0_i$ as well.  Although we have been unable to see how to do this with penalties, we have found empirically that it is unnecessary to impose any boundary conditions on this mode;  it is sufficient to enforce the condition on the incoming mode $U_-$.  

For a spacelike boundary with all characteristic modes outgoing, no boundary conditions and hence no penalties are required.  In that case, the boundary term in Eq.~\eqref{SDEdotSWC} is strictly non-positive.  On the other hand, if the boundary is spacelike with all characteristic modes incoming, a boundary condition can be enforced on $U_-$ \textit{and} $U_+$ by setting
\begin{flalign}
p\big|_{\d \Omega} &= \frac{1}{2}\Big(\,|\lambda_+|\,\delta U_+ - |\lambda_-|\,\delta U_-\Big), \label{p2}\\ \noalign{\vskip4pt}
q\big|_{\d \Omega} &= \frac{1}{2}\frac{|{\bf N}|}{\omega_x\,\omega_y\,\omega_z}\Big(\,|\lambda_+|\,\delta U_+ + |\lambda_-|\,\delta U_-\Big). \label{q2} 
\end{flalign}
In this case, the flux integrand in Eq.~\eqref{SDEdotSWC} would be
\beq
\begin{split}
F^n =& \, \lambda_- \big(U_-^\text{BC}{}^2 - \delta U_-^2 \big) + \lambda_+ \big( U_+^\text{BC}{}^2 - \delta U_+^2\big) \\
&\,+ 2 \,\lambda_0\, U^{0i}U^0_i. \label{FluxSWCsd2}
\end{split}
\eeq
In summary, the second-order system with penalties is
\begin{flalign}
\dot \psi =&\, -\alpha\pi + \ldots - \frac{|\lambda_-|}{2}\,\delta U_-, \label{SWCEq1} \\
\dot \pi =&\, -\alpha \gamma^{ij}\d_i\d_j\psi + \ldots + 
\frac{|\lambda_-|}{2}\frac{|{\bf N}|}{\omega_x\,\omega_y\,\omega_z}\,\delta U_-, \label{SWCEq2}
\end{flalign}
where as usual the penalties represent boundary values, the penalty on the first equation is applied throughout the interior of the domain via Eq.~\eqref{bulkp}, and the normal vector ${\bf N}$ is defined as in Eq.~\eqref{normal}.  Furthermore, it is to be understood that the penalties with $\delta U_-$ are applied only when $U_-$ is an incoming mode, and in the event that $U_+$ is also incoming, the penalties are modified according to Eqs.~\eqref{p2}-\eqref{q2}.

\subsection*{Numerical Tests}

\noindent In order to compute the error with respect to an analytical solution, we consider the inhomogeneous wave equation
\beq
\nabla_\mu\nabla^\mu \psi = \mathcal{S}, \label{SWCEqSource}
\eeq
where the source $\mathcal{S}$ is computed by substituting an analytical solution for $\psi$ into the left-hand side.  With a source term, Eq.~\eqref{CurvedEq1} remains unchanged, while Eq.~\eqref{CurvedEq2} is modified by adding a term $\alpha\,\mathcal{S}$:
\beq
\dot \pi = \ldots + \alpha\,\mathcal{S}. \label{SWCEq2Source}
\eeq
As an example of a test problem, we use the following background metric (the Schwarzschild solution in Kerr-Schild coordinates):
\beq
\begin{split}
ds^2 =& -\Big(1+\frac{2M}{r}\Big)\,dt^2 + \Big(1+\frac{2M}{r}\Big)\,dr^2 \\
&+ \frac{4M}{r}\,dr\,dt + r^2 d\Omega^2,
\end{split}
\eeq
and consider an analytical solution of the form
\beq
\psi_\text{analytical} = \cos(\omega\,t) e^{-(r-r_0)^2/\sigma^2} Y_{lm}. \label{soln}
\eeq
We evolve the second-order equations Eqs.~\eqref{SWCEq1}-\eqref{SWCEq2} with source term on a domain consisting of three concentric spherical shell subdomains.  The innermost boundary is placed just inside the horizon (located at $r=2M$), so that no boundary condition is required there (all characteristic modes are outgoing).  At interfaces between two subdomains, the value of the incoming mode $U_-^{\text{BC}}$ is supplied by the adjacent subdomain, while on the outermost boundary, $U_-^{\text{BC}}$ is computed from the analytical solution in Eq.~\eqref{soln}.  Time-stepping is performed using an explicit Runge-Kutta method.

\begin{figure}[t!]
\centering
\includegraphics[width=\columnwidth]{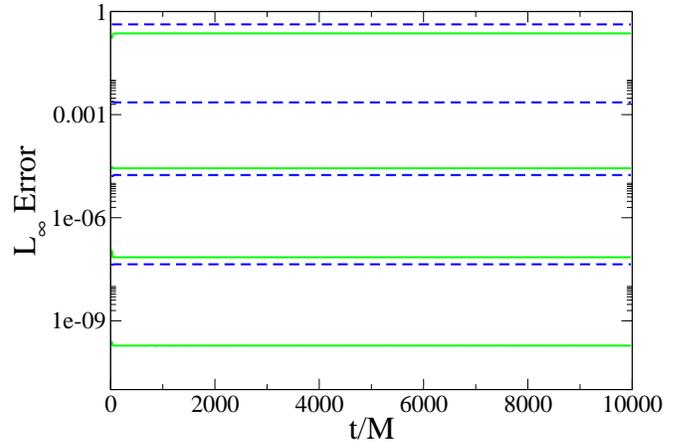}
\caption{$L_\infty$ error of a scalar wave given by Eq.~\eqref{soln} on a Schwarzschild background in Kerr-Schild coordinates evolved using the second-order system (solid green).  The results of equivalent first-order evolutions are plotted for comparison (dashed blue).  The domain consists of three concentric spherical shells with radial boundaries at $r=1.9, 11.9, 21.9, 32.9$ (in units of $M$).  The radial and angular resolutions $(N_r,L)$ of the runs are $(8,4)$, $(14,6)$, $(20,8)$, and $(26,10)$.  In this test, the following values for the analytical solution were used: $r_0=17\,M, \, \sigma=2\,M, \,\omega=0.5\,M^{-1}, \, l=3,$ and $m=1$.  The $L_\infty$ error is a moving average over an interval $\Delta t = 25$, which includes $50$ data points.}
\label{fg:SWCConv}
\end{figure}

For a spherical shell subdomain, we employ a spectral basis composed of Legendre polynomials in the radial direction scaled to the appropriate radial extent and spherical harmonics for the angular directions.  The numerical approximant for particular truncations $N_r$ and $L$ is therefore given by:
\beq
\psi = \sum\limits_{i=0}^{N_r} \sum\limits_{l=0}^L \sum\limits_{m=-l}^{+l} a_{ilm}\, \tilde P_i(r) Y_{lm}(\theta,\phi), \label{SphShellBasis}
\eeq
where $\tilde P_i(r)$ represents the appropriately scaled Legendre polynomial, and $a_{ilm}$ are the spectral coefficients.  Figure~\ref{fg:SWCConv} shows the $L_\infty$ error in $\psi$ for this test problem as a function of time for several resolutions.  For comparison, the results from an equivalent first-order evolution are plotted as well. 

%====================== FIGURE SWC test problem convergence ============================
%===========================================================================

The homogeneous second-order system of Eqs.~\eqref{SWCEq1}-\eqref{SWCEq2} is also stable and convergent in tests that we have run with arbitrary initial data on a variety of backgrounds.  For example, we have run simulations on a Schwarzschild background in Kerr-Schild, Painlev\'e-Gullstrand~\cite{MartelPoisson2001}, and fully harmonic coordinates, and on a Kerr background (with a spin up to $a=1$) in Kerr-Schild coordinates.  In those cases for which we do not have an analytical solution to supply a condition on $U_-$ at the outer boundary, we find good results by using a Sommerfeld condition, assuming a solution of the form
\beq
\psi \sim \frac{f(t-r)}{r}.
\eeq
This translates into a condition on the incoming mode at the outer boundary:
\beq
U_-^\text{BC} \sim \frac{\psi}{r}.
\eeq

For the first-order system, as discussed below Eq.~\eqref{1dModIP}, one can use Chebyshev polynomials instead of Legendre for the radial basis in Eq.~\eqref{SphShellBasis}, and the results are comparable (the same holds for Einstein's equations as well).  In the second-order evolutions, while it is acceptable to use a Chebyshev basis for flat space applications, we find that the error is significantly larger (almost two orders of magnitude) in the curved background case.  Note, however, that we are not addressing the Chebyshev-Legendre method discussed below Eq.~\eqref{1dModIP} here, but rather the use of a Chebyshev basis without modifying the index values of the penalty functions.  Presumably, the Chebyshev-Legendre method would perform equally well, but we have not explored this modification.

\section{Discussion}

We have not found a significant difference in efficiency between the first- and second-order forms of the equations for the simple systems considered in this paper.  Even with equal time steps, the rates of the first- and second-order codes (using explicit time-stepping) are comparable (within about 10\% of each other), despite the absence of the third set of evolution equations (the $\dot \phi_i$ equations) in the second-order system.  One reason for this is that second derivatives are computationally more expensive than first derivatives on arbitrary domains, because of the coordinate transformations involved.  Symbolically, the transformation of a first derivative to new (barred) coordinates involves a multiplication by the Jacobian $J$ of the transformation:
\beq
\bar \d \psi = J \d \psi,
\eeq
whereas the transformation of a second derivative requires the Hessian $H$ and the first derivative as well:
\beq
\bar \d^2 \psi = JJ \d^2 \psi + H \d \psi.
\eeq

It is evident that the second-order evolutions generally have smaller errors than their first-order counterparts at given resolutions by as much as two orders of magnitude, as can be seen in Figs.~\ref{fg:FlatConv}-\ref{fg:SWCConv}.  To a degree, then, this makes the second-order evolutions somewhat more efficient in the sense that for a given accuracy goal, a smaller resolution is required in the second-order system.  However, our focus has not been on comparing or analyzing code efficiencies, but rather on establishing the viability of a second order in space spectral method.  We believe the second-order system has the potential to show substantial increases in efficiency for more complicated systems than those considered here.

All of the energy arguments in this paper have assumed a grid structure that can be mapped to a cube.  While it is straightforward to apply these methods to spherical or cylindrical shells, it is assumed that any dimension with boundaries (e.g.\ the radial dimension on a spherical shell) has a collocation grid that contains its endpoints (Gauss-Lobatto grid).  For a domain containing the origin, such as the unit disk, it is typical to use a radial grid of Gauss-Radau points, so that the endpoint at the origin is omitted (see e.g.\ Ref.~\cite{MatsushimaMarcus1995}).  We have not considered the generalization to such domains.

We have shown how to evolve a multi-domain second order in space wave equation stably using spectral methods.  For more general systems, energy arguments like those given in this paper cannot be carried out.  Nevertheless, in the most important case we consider, namely Einstein's equations in generalized harmonic form, these methods work well.  The reason is that the \textit{principal part} of the equations is directly analogous to the scalar wave equation on a curved background~\cite{Lindblom2006}.  We will report on these extensions in a subsequent paper~\cite{Taylor2010}.

\begin{acknowledgments}
We would like to thank Carsten Gundlach, Jan Hesthaven, and Manuel Tiglio for valuable comments.  Most of the numerical tests were performed within the framework of the Spectral Einstein Code (SpEC) developed at Cornell and Caltech primarily by L.K., Harald Pfeiffer, and Mark Scheel~\cite{SpECwebsite2010}.  This work is supported in part by grants from the Sherman Fairchild Foundation to Caltech and Cornell, and from the Brinson Foundation to Caltech; by NSF grants PHY-0601459, PHY-0652995, and DMS-0553302 at Caltech; by NSF grants PHY-0652952, DMS-0553677, PHY-0652929, and NASA grant NNX09AF96G at Cornell.
\end{acknowledgments}

\appendix

\section{Gauss-Legendre-Lobatto Quadrature} \label{appGLLnotes}

\noindent Here we provide some of the properties of Gauss-Legendre-Lobatto quadrature~\cite{Boyd2001}.  The basis functions on $[-1,1]$ are the Legendre polynomials $P_n(x)$.  This is a convenient choice for obtaining analytical results because the Legendre polynomials are orthogonal under a weighting function of unity:
\beq
\int\limits_{-1}^{1} \rho(x) P_n(x)\,P_m(x)\,dx = \frac{2}{2n+1}\,\delta_{nm},
\eeq
with $\rho(x)=1$.  The $(N+1)$-point quadrature rule,
\beq
\int\limits_{-1}^1 u(x)\,dx \simeq \sum\limits_{i=0}^N \omega_i\,u(x_i),
\eeq
is exact if $u(x)$ is a polynomial of degree $2N\!-\!1$ or less.  The $N\!+\!1$ nodes $x_i$ are 
\begin{flalign}
x_0 = & -1, \\
x_N = & +1, \\
x_i = & \> \text{the roots of}\, P_N^\prime(x) \\
&\> \text{for}\, 0 < i < N,
\end{flalign}
and the weights $\omega_i$ are given by
\beq
\omega_i = \frac{2}{N(N+1)[P_N(x_i)]^2}. 
\eeq
Note that there is no known explicit formula for the roots of $P^\prime_N(x)$---they must be found numerically.  A function $\psi(x)$ is approximated by an $N^\text{th}$-order interpolating polynomial $\psi_N(x)$, which can be expressed as
\beq
\psi_N(x) = \sum_{i=0}^N \psi(x_i) C_i(x),
\eeq
where $C_i(x)$ are cardinal functions satisfying $C_i(x_j) = \delta_{ij}$.  They can be written as
\beq
C_i(x) = \frac{-(1-x^2)P^\prime_N(x)}{N(N+1)P_N(x_i)(x-x_i)}.
\eeq
Differentiation can be computed via matrix multiplication from
\beq
\psi^\prime_N(x_i) = \sum_{j=0}^N D^{(1)}_{ij} \psi(x_j),
\eeq
where $D^{(1)}_{ij} \equiv C_j^\prime(x_i)$ is the first-order differentiation matrix.  The second-derivative matrix is defined similarly and satisfies $D^{(2)} = D^{(1)}D^{(1)}.$  An efficient algorithm for computing pseudo-spectral differentiation matrices is given in Ref.~\cite{Fornberg1998}.

If $f,g$ are two $N^\text{th}$-order polynomials, summation by parts follows naturally because the product $fg^\prime$ is a polynomial of order $2N\!-\!1$ or less:
\beq
\langle f, g^\prime \rangle = \sum\limits_{i=0}^N \omega_if_ig^\prime_i
= f_i\,g_i \big|_{i=0}^N - \langle f^\prime, g \rangle.
\eeq
Summation by parts generalizes to higher dimensional inner products in a straightforward way.  For example, if $f$ and $g$ are $2$-d polynomials in $x$ and $y$:
\begin{flalign}
\langle \d_xf, g \rangle = & \sum\limits_{i,j=0}^{N} \omega_i \omega_j (\d_xf)_{ij}\, g_{ij} \\
= & \sum\limits_{j=0}^{N} \omega_j (fg)\big|_{i=0}^N - \langle f, 
\d_x g \rangle.
\end{flalign}

\section{Proof of Inability to Generalize 1D Penalty Function} \label{appDOFProof}
\noindent In this section we will show that in two or more dimensions the inner product 
\beq
\langle \d^i\d_i \psi, p \rangle
\eeq
that arises in the energy arguments discussed in this paper cannot be made to vanish in general with a penalty function $p$ that satisfies the boundary conditions.  We will argue by counting degrees of freedom.  For simplicity, consider the two-dimensional case and let the domain be a square with $N\!+\!1$ grid points along each dimension.

Instead of using a basis of Legendre polynomials, consider a (non-orthogonal) basis of functions $x^iy^j$.  A scalar field is thus approximated on the grid as a two-dimensional interpolating polynomial of the form
\beq
\psi = \sum\limits_{0 \le i,j \le N} a_{ij} x^i y^j. \label{psiExpan}
\eeq  
There are $(N\!+\!1)^2$ basis functions and hence the same number of degrees of freedom in the function $\psi$.  The penalty function $p$ must satisfy $4N$ boundary conditions on the square.

Now consider operating on the expansion of $\psi$ in Eq.~\eqref{psiExpan} with the Laplacian $\d^2_x + \d^2_y$.  The effect of this operation on a term $x^i y^j$ is essentially
\beq
x^i y^j \rightarrow x^{i-2}y^j + x^i y^{j-2}. \label{xyterm}
\eeq
Since we are only interested in counting the degrees of freedom that remain in $\nabla^2 \psi$, we only need to retain one of the terms in Eq.~\eqref{xyterm}:
\beq
x^i y^j \rightarrow x^{i-2}y^j. \label{xyterm2}
\eeq
By doing this, we will at worst undercount the degrees of freedom in $\nabla^2 \psi$.  This leaves terms of the form $x^{i-2}y^j$ for $2 \le i \le N$ and $0 \le j \le N$, which implies that there are at least $(N\!+\!1)(N\!-\!1)$ degrees of freedom remaining in the Laplacian.

There are thus at most $(N\!+\!1)^2 - (N\!+\!1)(N\!-\!1) = 2N\!+\!2$ degrees of freedom for constructing a penalty function that is orthogonal to $\nabla^2 \psi$ (for arbitrary $\psi$), which is not enough to satisfy the $4N$ boundary conditions.  The same argument can be applied in any number of dimensions.  In particular, in the three-dimensional case we find that there are at most $2(N\!+\!1)^2$ degrees of freedom for constructing the penalty function---not enough to satisfy the $6N^2\!+\!2$ boundary conditions, which proves the assertion made below Eq.~\eqref{3dedot}.

\section{Derivation of 3D Penalty} \label{appBulkPenalty}

\noindent In this section the form of the three-dimensional bulk penalty given by Eq.~\eqref{bulkp} will be derived.  The goal is to minimize the inner product
\beq
\langle \d^i\d_i \psi, p \rangle,
\eeq
with the values of the penalty function $p$ on the boundary given.  First, we will revisit the one-dimensional problem on the interval $[-1,1]$ from a new point of view.  In Section~\ref{MainResult} it was shown that the one-dimensional inner product $\langle \psi^{\prime\prime},p \rangle$ vanishes when $p$ is constructed from the functions $f,g$ defined in Eqs.~\eqref{fdef}-\eqref{gdef}.  Recall that the functions $f$ and $g$ were constructed from $P_N$ and $P_{N-1}$ to be orthogonal to $\psi^{\prime \prime}$.  

Let us start over and consider the one-dimensional penalty function $p$ to be unspecified, except on the boundaries.  Suppose also that there is no boundary condition at $x_N=+1$, so the penalty function satisfies $p_N=0$.  The boundary condition at $x_0=-1$ fixes the value $p_0$, and we can view the values $p_i$ for $0 < i < N$ as free parameters for minimizing the inner product:
\beq
\langle \psi^{\prime \prime}, p \rangle = \omega_0 \psi^{\prime \prime}_0\, p_0 + \sum\limits_{i=1}^{N-1} \omega_i \psi^{\prime \prime}_i\, p_i.  \label{pvanish1}
\eeq
This will vanish if and only if
\beq
\psi^{\prime \prime}_0 = \sum\limits_{i=1}^{N-1} \Big(\frac{-\omega_i p_i}{\omega_0 p_0}\Big) \psi^{\prime \prime}_i,  \label{pvanish2}
\eeq
where it is safe to assume $p_0 \neq 0$ (if $p_0 = 0$, then $p=0$ as there would be no need for a penalty function).  Since $\psi^{\prime \prime}(x)$ is an {\em arbitrary} $(N\!-\!2)$-order polynomial, this equation defines the ideal interpolation weights $c_0(x_i)$ for approximating a function at $x_0$~$=$~$-1$ based on its values over a stencil of points $x_i$ for $0 < i < N$:
\beq
\psi^{\prime \prime}_0 = \sum\limits_{i=1}^{N-1} c_0(x_i) \psi^{\prime \prime}_i.
\eeq
Assuming the grid points are Gauss-Legendre-Lobatto points, we can therefore make the identification
\beq
c_0(x_i) = -\frac{\omega_i p_i}{\omega_0 p_0} = -\frac{\omega_i}{\omega_0} f_i, \label{c0def}
\eeq
where $f_i$ are the grid values for $0<i<N$ of the function $f$ defined in Eq.~\eqref{fdef}, and we have used the fact that Eq.~\eqref{pvanish2} holds when the penalty $p$ is defined by Eq.~\eqref{pdef}.

The case with a boundary condition at $x=+1$ and not at $x=-1$ (so $p_0=0$) is similar, and we conclude that the interpolation weights $c_N(x_i)$ for approximating a function at $x_N=+1$ based on its values at points $x_i$ for $0<i<N$ are given by
\beq
c_N(x_i) = -\frac{\omega_i}{\omega_N} g_i, \label{cNdef}
\eeq
where $g_i$ are the grid values for $0<i<N$ of the function $g$ defined in Eq.~\eqref{gdef}.

Now let us consider the two-dimensional problem on the square $[-1,1]\times[-1,1]$.  In the following we will use the index $i$ exclusively for summing over $x$ values and $j$ for $y$.  Our goal is to construct the values of $p$ on the interior of the domain so as to minimize the inner product
\beq
\langle \Delta \psi, p \rangle = \sum\limits_{ij}\omega_i\omega_j \Delta\psi_{ij} p_{ij}, \label{IP}
\eeq
where $\Delta$ represents the Laplacian operator $\d^i\d_i$, and we consider the values of $p$ on the boundary to be given.  Consider a point on the edge at $(x_0, y_j)$, for example.  The term in the inner product due to this point is
\beq 
\omega_0 \omega_j \Delta \psi_{0j} p_{0j}. \label{edgept}
\eeq
Now define $p$ on the interior along the $j^{th}$ row to be 
\beq
p_{ij} = p_{0j} f_i, 
\eeq
just as in the one-dimensional case.  Using the identification of $f$ as interpolation weights from Eq.~\eqref{c0def}, the contribution to the inner product from the interior of this row is
\begin{flalign}
\langle \Delta \psi, p\rangle\big|_{j^\text{th} \text{row}} =& 
\sum\limits_{i=1}^{N-1} \omega_i \omega_j \Delta \psi_{ij} p_{ij} \\ 
=& - \omega_0 \omega_j p_{0j} \sum\limits_{i=1}^{N-1} \Delta \psi_{ij} \,c_{0i} \label{c0i}\\ \noalign{\vskip 3pt}
\simeq & - \omega_0 \omega_j \Delta \psi_{0j} p_{0j}, 
\end{flalign}
which approximately cancels the term from the point on the edge in Eq.~\eqref{edgept}.  In Eq.~\eqref{c0i} we have written $c_{0i}$ for the interpolation weights $c_0(x_i)$ defined in Eq.~\eqref{c0def}.  Next, consider a point at a corner, say $(x_N, y_0)$.  The term in the inner product due to this point is 
\beq
\omega_N \omega_0 \Delta \psi_{N0} \,p_{N0}. \label{cornerpt}
\eeq
Define $p$ on the interior of the domain to be 
\beq
p_{ij} = - p_{N0} \,g_i f_j.
\eeq
The contribution of $p_{ij}$ to the inner product on the interior of the square is now
\begin{flalign}
\langle \Delta \psi, p \rangle\big|_{\text{interior}} =& 
\sum\limits_{i,j=1}^{N-1}\omega_i \omega_j \Delta \psi_{ij}\, p_{ij} \\ 
=& - \omega_N \omega_0 \,p_{N0} \sum\limits_{i,j=1}^{N-1}\Delta \psi_{ij} \,c_{Ni} c_{0j} \\ \noalign{\vskip 3pt}
\simeq &  - \omega_N \omega_0 \Delta \psi_{N0}\, p_{N0}, 
\end{flalign}
which approximately cancels the contribution from the point on the corner in Eq.~\eqref{cornerpt}.  Following this procedure, we construct $p$ on the interior by adding a contribution from each boundary segment: $4$ edges and $4$ corners on this $2$-d domain.  Explicitly, this gives
\beq
\begin{split}
p_{ij} = & \, p_{0j}f_i + p_{N\!j}\,g_i + p_{i0}f_j + p_{iN}\, g_j \\
& - p_{00}f_i f_j - p_{0N} f_i g_j - p_{N0}\, g_i f_j \label{2dp} \\
&  - p_{N\!N} \,g_i g_j. 
\end{split}
\eeq
This generalizes to three or more dimensions in a straightforward way.  Each term in $p_{ij}$ has a number of products of $f$ or $g$ equal to the codimension of the boundary segment it depends on.  The only caveat is that the sign of the terms added to $p$ should be $(-1)^{m+1}$, where $m$ is the codimension of the boundary piece producing the term.  This is evident in the $2$-d example above where the terms in Eq.~\eqref{2dp} due to the corners are negative.  The sign difference arises simply because of the negative sign in the relation between the interpolation weights $c_0,c_N$ and the functions $f,g$ in Eqs.~\eqref{c0def}-\eqref{cNdef}.

With the penalty function $p$ constructed according to the above procedure, the inner product of $p$ with \textit{any} analytic function $h$ (hence $\Delta \psi$) satisfies
\beq
\langle h, p \rangle \, \rightarrow \, 0, \quad \text{as} \quad N \rightarrow \infty. \label{hinprod}
\eeq
In particular, we have shown that the last term in Eq.~\eqref{3dedot} asymptotically vanishes, as claimed below Eq.~\eqref{bulkp}.  Moreover, while we have not bounded the error for a given resolution, the inner product in Eq.~\eqref{hinprod} will be as small as possible in the sense that it vanishes for the polynomial approximations to $h$ up to order $N-2$.

\end{document}